# Quantitative phase imaging of retinal cells

Timothé Laforest[1*a], Dino Carpentras[1*], Laura Kowalczuk[2], Francine Behar-Cohen[2], Christophe Moser[1]

1 School of Engineering, LAPD, École Polytechnique Fédérale de Lausanne (EPFL), Lausanne, Switzerland

2 University of Lausanne, Department of Ophthalmology, Lausanne, Switzerland

*authors contributed equally to this work
a: timothe.laforest@epfl.ch

**Abstract**

The health of cells found in the inner retinal layers of the human eye is crucial to understand the onset of diseases of the retina such as macular degeneration and retinopathy. A challenge is to periodically image these cells in human eyes to detect abnormalities well before physiological and pathological changes occur. However, *in vivo* imaging of many of these cells is still elusive despite the phenomenal advances in Optical Coherence Tomography (OCT) and Adaptive Optics systems. It stems from the fact that cell contrast in reflection is extremely low. Here, we report a major advance towards this goal by proposing and demonstrating a method to visualize these cells with high contrast and resolution. The method uses a transcleral illumination which provides a high numerical aperture in a dark field configuration. The light backscattered by the Retinal Pigment Epithelium (RPE) and Choroid layer provides a forward illumination for the upper layer of the retina, thus providing a transmission illumination condition. By collecting the scattered light through the pupil, the partially coherent illumination produces dark field images, which are combined to reconstruct a quantitative phase image with twice the numerical aperture given by the eye's pupil. The retrieved quantitative phase images (QPI) of cells in *ex vivo* human and pig's eyes are validated with those taken with a standard

QPI system. We then report, to our knowledge, the very first human *in vivo* phase images of inner retinal cells with high contrast.

# Introduction

The recent developments in retinal imaging have contributed to the elucidation of disease mechanisms, permitted treatment effects validation and guided treatments regimen in clinical practice. The eye is a direct extension of the central nervous system and thus offers a unique opportunity to directly observe ocular structures through the clear and transparent ocular tissues and media. Since ophthalmoscopy, numerous methods have been developed to visualize structures in the retina with enhanced definition. Until now, every optical method developed to observe the interior of the human eye uses the information contained in the reflected light to reconstruct an image of the interior structure. The photons entering the eye go through the cornea, the lens, the ocular aqueous and gel media before reaching the retina, where the photons must go through several layers of neurons and interneurons before reaching the photoreceptor cells where phototransduction takes place. Human retinal thickness ranges from 150 to 300 um depending on the area of the retina, the gender and pathology of the subject [1]. The retina is maintained in a relative dehydrated state to allow light transmission through the translucent layers. As a consequence, the retina has very low reflectivity. Despite this, Optical Coherence Tomography/Microscopy is a very successful method commonly used in routine eye examination to observe retinal layers by interfering a reference beam with the beam reflected by the slight index of refraction mismatch in the stratified structure of the retina [2,3]. Linear confocal imaging with adaptive optics is also routinely used to resolve cone cells in the fovea because the later are highly reflective [4].

Recent works proposed optical methods to observe cells and microvasculature with phase contrast in the retina. Toco et al used confocal dark field imaging of microvasculature close to the optic disc thanks to an offset aperture through the pupil [7-8]. Another study reported results based on split-detector, combined with an adaptive optics scanning laser ophthalmoscope (AOSLO) [9-11]. This approach produces phase contrast images of ganglion cells in the mouse retina or microvasculature by placing intensity masks in the retinal conjugate plane in front of the AOSLO point detector. The split-detector allows obtaining differential measurements, thus removing background and absorption terms in the image. However, the contrast of phase objects is limited due to transpupillary illumination which produces specular light, and the laser scanning modality suffers from a relatively low frame rate compared to an en face camera based system.

Despite these tremendous developments in imaging techniques, nearly half of the retinal structures can still not be imaged in vivo [25].

Recently, other techniques for phase imaging have been developed outside the field of ophthalmology. Mertz et al developed a method for phase contrast imaging in reflection using the scattering properties of biological tissues [13]. This method takes advantage of the scattering properties of the deep layers of the object to provide oblique illumination to the superficial layers. By recording two images of the sample with different oblique illumination, one can reconstruct a phase gradient image and perform a volumetric scan of a sample up to 100 µm thick [14].

Another phase contrast method uses a transmission system and several oblique illuminations to obtain quantitative phase information [15-16]. This method, developed by Waller and co-workers, is based on a differential illumination of the sample. Recording two images with asymmetric illumination allows computing the differential phase contrast image (DPC) which highlights the edges of the phase objects in the sample while removing the absorption features [17-18]. A phase retrieval algorithm can then be applied assuming a weak object in phase and amplitude to obtain a quantitative phase image. Quantitative phase measurement of thin slices allows a broad range of characterization of cells, such as cytotoxicity screening [30] or dynamic cell morphometry [31].

More generally, phase imaging techniques such as differential interference contrast [19-20] and digital holographic (DHM) [21] microscopies are limited to a transmission arrangement, which has thus far prevented its applicability to *in vivo* ophthalmology.

Here, we present a method that circumvents this limitation and gives high resolution quantitative phase imaging of the retina. Our approach consists in using transcleral illumination with multiple oblique directions to record dark field images of the retina, and use them to reconstruct a quantitative phase image. Transcleral illumination has been used for decades for diagnostic in oncology of the eye but never for cell imaging. By passing through the sclera, light produces a high angle oblique illumination of the retina, which provides a larger illumination numerical aperture than the collecting aperture of the pupil. Hence, transcleral illumination with a partially coherent LED light allows for super-resolution [16]. The processing of two asymmetric dark field images removes the absorption and background terms of the collected beam by computing the differential phase contrast image. Knowing the illumination transfer function, it is possible to recover the quantitative phase information of the sample. Since back scattered light is used as an effective transmission illumination, an accurate modeling of the illumination diffusion is also required. To reach in vivo cellular resolution in the retina, one must undo the optical aberration of the eye. Several aberration correction methods exist. A first method relies on using adaptive optics, resulting in an accurate but complex setup [22]. A second option is to correct the aberration computationally using a guide star algorithm [23] or a sharpness based criteria algorithm [24].

The paper is organized as follows: in section 1 we introduce the method and the system. We first assess the quantitative phase measurement of the proposed technique in an *ex vivo* apparatus mimicking the apparatus used for *in vivo* imaging. We verified the doubling of resolution compared to coherent illumination using the USAF intensity test target. Next, we compare our modality to digital holographic microscope images to show the quantitative phase measurement. We also compare a phase image with confocal intensity reflection images to demonstrate the improvement of contrast for phase features. We then present experimental 3D results of the proposed phase imaging technique using ex-vivo samples of human and pig eyes. Finally, we report in vivo phase images of the human eye, without any adaptive optics and wavefront correction, using post processing for aberration reduction [24,29].

**Detailed description of the method**

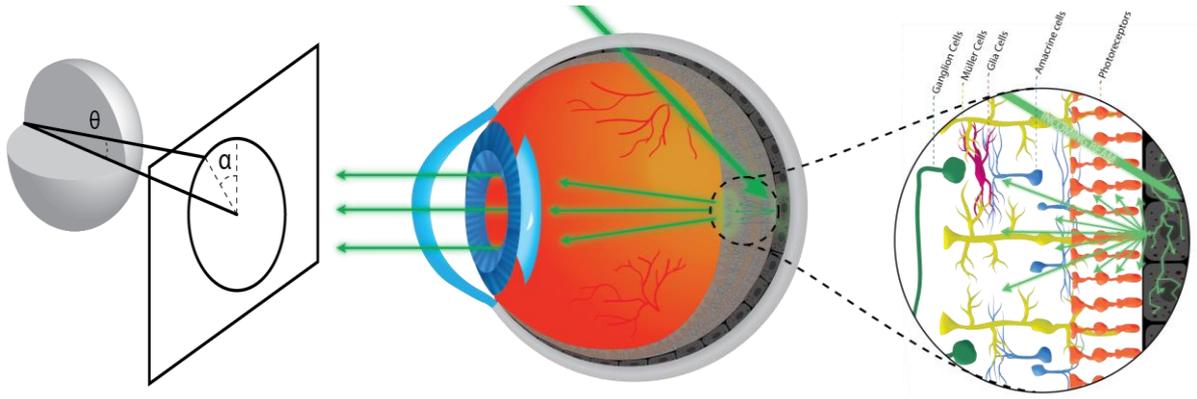

Fig. 1 Illumination of the retinal layers provided by transcleral illumination. The light is first transmitted through sclera, RPE and retina. After travelling through the aqueous humor it impinges on the RPE. Here backscattering off the RPE generates a new illumination beam. This secondary illumination provides a transmission light propagating through the translucent layers of the retina which is then collected by the pupil. Azimuthal angle θ and polar angle ɑ.

Transcleral illumination of the fundus has a higher numerical aperture than what is obtainable via illumination through the pupil. In Fourier domain, this is equivalent to a shift of the illumination towards higher spatial frequencies, meaning exciting the highest spatial frequencies. In addition, oblique illumination coupled with an imaging system which captures images through the pupil, but does not collect the superficial specular reflection, produces dark field images of the fundus, thus providing a high Signal to Noise ratio (SNR) allowing to detect the dim light reflected from the fundus deep layers. The forward scattered light by the sclera also illuminates the fundus uniformly. Additionally, the use of spatially incoherent light gives a system's bandwidth that is twice the bandwidth obtainable with coherent light. A dark field image can then be obtained with just one illumination source. Phase imaging is performed thanks to the following procedure: at least two pictures, I(ɑ) and I(ɑ+180) captured with two different asymmetric illumination polar angles ɑ and ɑ+180° are required. One can then obtain the differential phase contrast (DPC) image according to:

$$I_{DPC} = \frac{I(\alpha) - I(\alpha+180°)}{I(\alpha) + I(\alpha+180°)}$$

By combining different DPC images of the same target, it is possible to obtain a quantitative phase image [15]. The illumination direction is an important parameter in the phase reconstruction process.

**Ex-vivo demonstration**

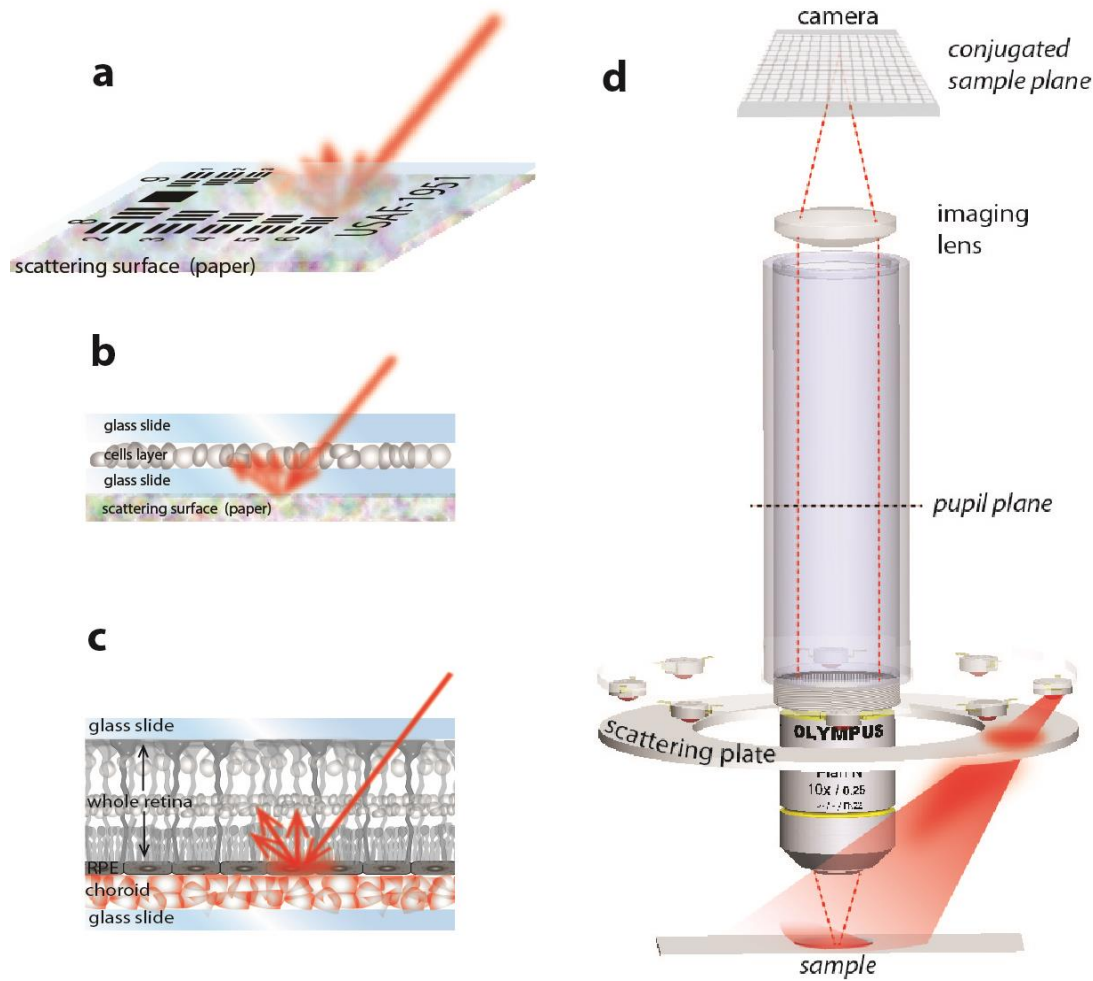

Figure 2: a. Intensity USAF target sample mounted on top of a scattering surface (paper) for assessing the modulation transfer function of the optical system. Incoming light back scatters from the paper.b. Flat mounted 10 um thick slice of pig retinal cells. A scattering surface (paper) is attached to the bottom glass slide in order to provide back scattering light.. c. Flat-mounted sample of a fixed retina with the RPE and choroid. Incoming light back scatters from the RPE-choroid d. Experimental setup used. LED light illuminates the samples described in a)-c). The backscattered light effectively becomes a transmitted beam that illuminates the transparent phase layers before reaching the microscope objective. An imaging lens makes an image of the sample on a camera.

The experimental setup uses a 0.25 NA objective that matches approximately the maximum 0.24 NA of a fully dilated human eye. The illumination lights are 650 nm LEDs with 50 nm bandwidth. The scattering of the sclera is simulated by placing a scattering paper plate between the LEDs and the samples as illustrated on Fig 2.d. We first measured the modulation transfer function (MTF) of our system using the groups 8 and 9 of a USAF intensity target. Fig. 3 a shows that the measurement fits well the ideal MTF of an incoherent illumination having a cutoff frequency of 2NA/lambda. It allows to resolve the 780 nm wide bars with a 0.25 NA objective and 650 nm wavelength light thus confirming the doubling of the resolution.

Then, we perform imaging of an ex vivo pig retina. The retina sample was prepared by slicing 10 um thick slices using a cryosection process, and flat mounted on a coverslip. A scattering surface (paper) has been attached to the bottom glass slide in order to simulate the back scattering light, as shown in Fig 2.b. The 10 um thick horizontal sections of the retina allow imaging with both a digital holographic microscope (DHM) and the proposed method. A DHM produces true phase images, which allows a quantitative comparison with the proposed method. The images from the DHM are taken with a 0.4 NA microscope objective. The proposed method uses a 0.25 NA objective to match the maximum NA of the eye. As demonstrated above, incoherent illumination allows doubling of the resolution compared to coherent illumination. Hence the effective NA of the proposed method is 0.5 and thus produces higher resolution images than the 0.4 NA DHM image, which uses coherent illumination.

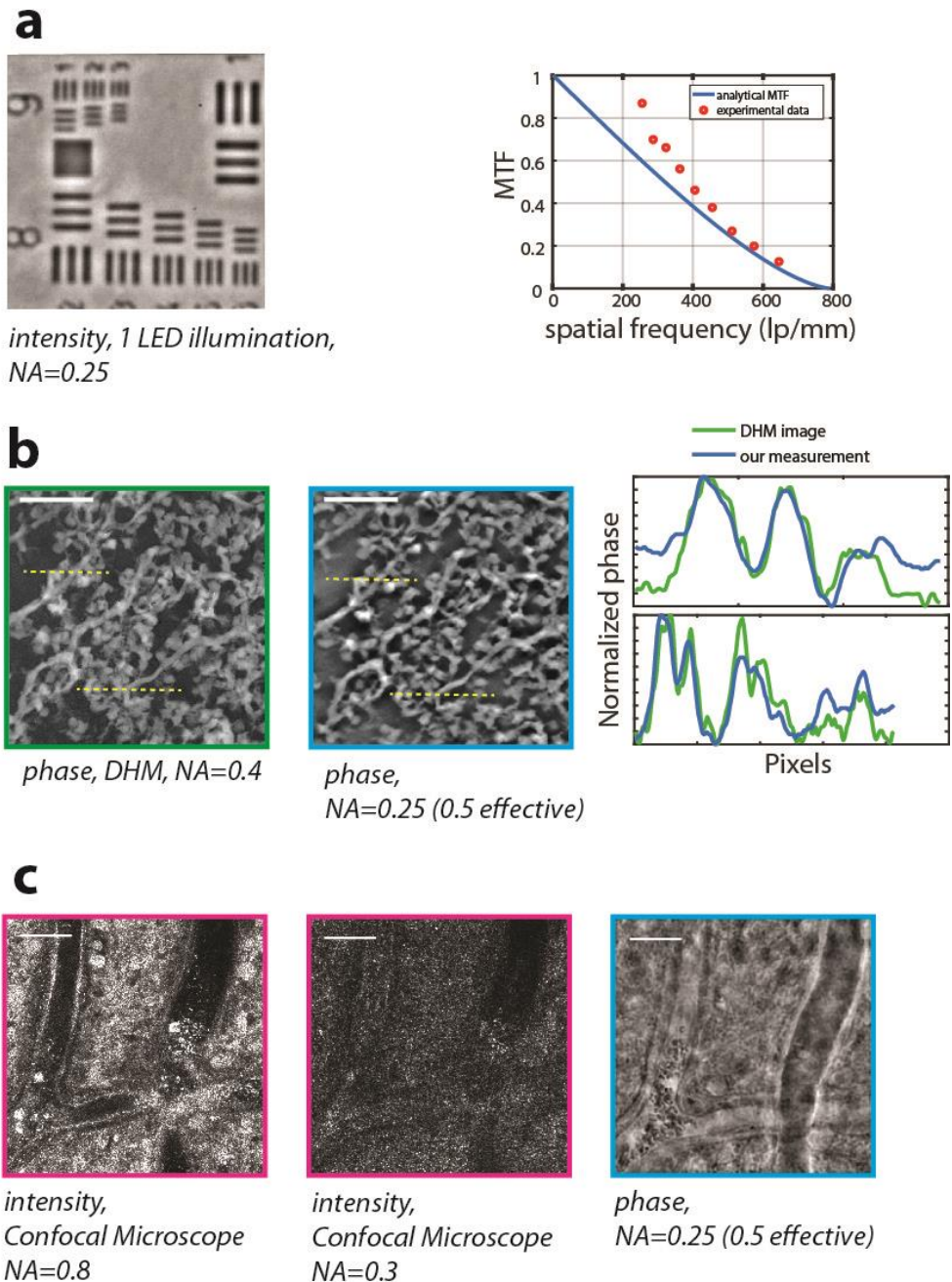

Figure 3 : a. One LED illumination of the USAF intensity target and plot of the theoretical and experimental modulation transfer function from the image. Red dots correspond to the contrasts of elements of groups 8 and 9 of the USAF. b Phase image of a 10 um thick slice of pig retina inner nuclear layer. Comparison with the proposed phase imaging method in reflection (center) and DHM phase imaging (left). Note the artifacts in the DHM images due to Fourier filtering of the holograms. Scale bars = 30 um. Horizontal cross sections of the images (right). C. Image of a flat-mounted human choroid and retina. Comparison of confocal reflection images with 0.8 NA (left) and 0.3 NA (center

column) numerical aperture and phase image with 0.25 NA, corresponding to 0.5 effective NA (right). Scale bars = 50 um.

Furthermore, we compare our phase imaging modality to a reflection confocal microscopy (Zeiss). A sample of human retina was fixed and flat mounted with the choroid, as shown in Fig. 2.c, the back illumination being provided by the scattering of the RPE-choroid layers. In order to compare the images, the region was chosen on vessels. Fig 3c. shows the intensity images taken with 0.8 NA (left) and 0.3 NA (center) confocal microscope. The phase image (right) recorded at the same depth exhibits better SNR and resolution compared to the 0.3 NA confocal microscope image. In addition, compared to the 0.8 NA image, the phase image makes visible other features.

Finally, we demonstrate depth phase images of a pig sample of the whole retina and choroid (see Fig 2.c). Here, the back illumination is also provided by the scattering of the RPE-choroid layers. Depth scanning is obtained by moving the sample in z. The 0.25 NA objective used in the experimental setup gives a depth of field of 8.5 μm, which allows imaging every layer. The imaged region on the retina is the 'area centralis' [24]. Fig. 4 shows the resulting phase images which are digitally tagged to highlight the cells nuclear and density of every layer. Figures 4a and 4b represent the nerve fiber layer (NFL), figures 4c and 4d the ganglion cells layer (GCL), figures 4e and 4f the inner plexiform layer (IPL), showing the microglia cells (red arrows), figures 4g and 4h the inner nuclear layer (INL), figures 4i and 4j the outer nuclear layer (ONL), and figures 4k and 4l the photoreceptor layer (PR). The raw images has been process in order to tag the cells nuclear, and compute the cells density.

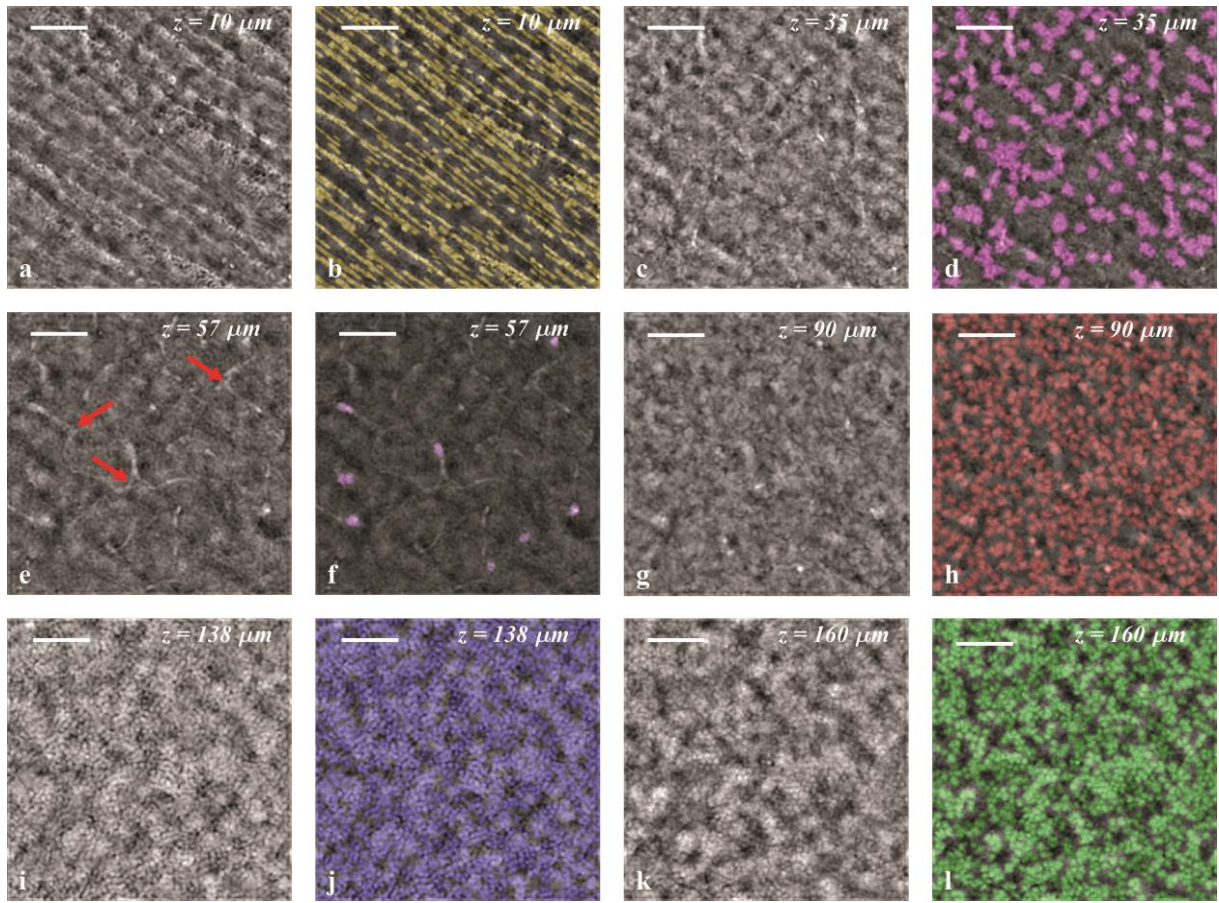

Figure 4 : Scan in depth of flat-mounted choroid and retina of a pig's eye. a, c, e, g, i, k: raw phase images. b, d, f, h, j, l: phase images with digitally tagged cells and structures. Nerve fiber layer (z = 10 um) a,b. Ganglion cells layer (z = 35 um) c,d. Inner plexiform layer, with microglia (red arrows) (z = 57 um) e, f. Inner nuclear layer (z = 90 um) g,h. Outer nuclear layer (z = 138 um) i, j. Photoreceptors layer (z = 160 um) k, l. Scale bars = 50 um.

The density analysis for the GCL, INL, ONL and PR layer is reported on figure 5. The data from the literature of pig retinal cells density were compared to our experimental density values for the GCL [26] and photoreceptors layer [27, 28]. In [26] the density range of the ganglion cells is 1500 to 4000 cells/mm² while we measured 2260 cells/mm² on our sample. The cones densities given in references [27] and [28] are respectively 19000 to 22600 cells/mm² and 18000 to 21500 cells/mm² and we obtained in average 20930 cells/mm².

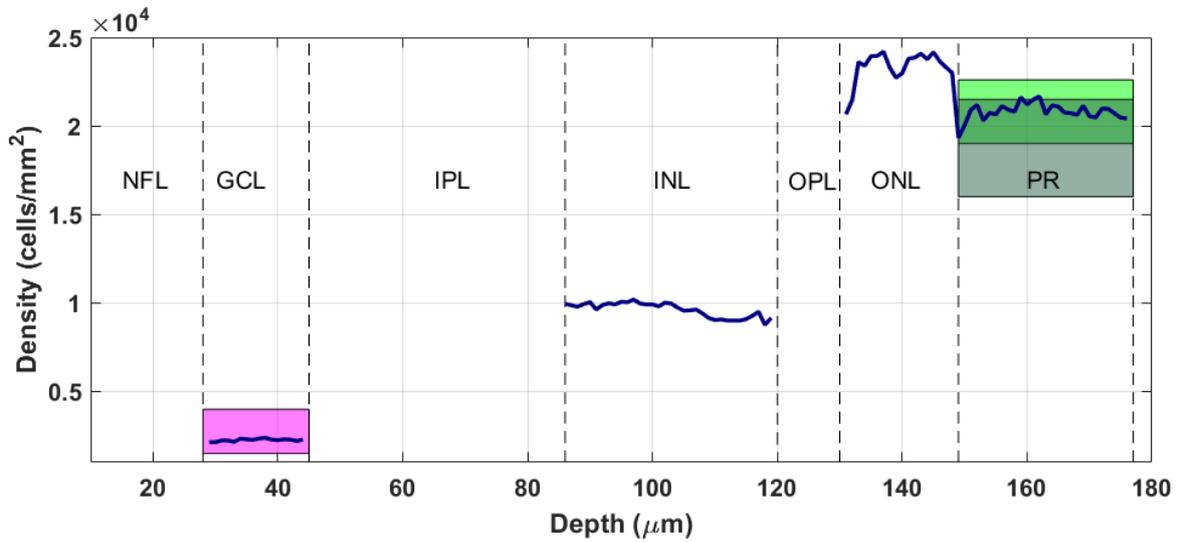

Figure 5 : Cells density over depth. Magenta: density range of ganglions from [26]. Green : density range of cones [27]. Dark green: density range of cones from [28].

**In-vivo demonstration**

For in vivo demonstration, in order to produce angled illumination on the retina, we built custom flexible PCBs on which small footprint LEDs were soldered, as shown in Fig 6a and 6b. Next, the different LEDs were synchronized to a sensitive EMCCD camera thanks to a programmable board driving the LEDs and generating the camera input trigger (see Fig 6d). Fig 6d illustrates the optical setup. It consists of 3 lenses and a badal system in order to correct for eye defocusing. When the badal system is set at the appropriate position, a point source at the retina creates an image point on the camera with a magnification equal to f1f3/feyef2, where feye, f1, f2, f3 are the focal distances of the eye lens, lens L1, lens L2 and lens L3 respectively. A diaphragm (D) at the pupil plane allows filtering the back scattered parasitic light from the PCB or the skin.

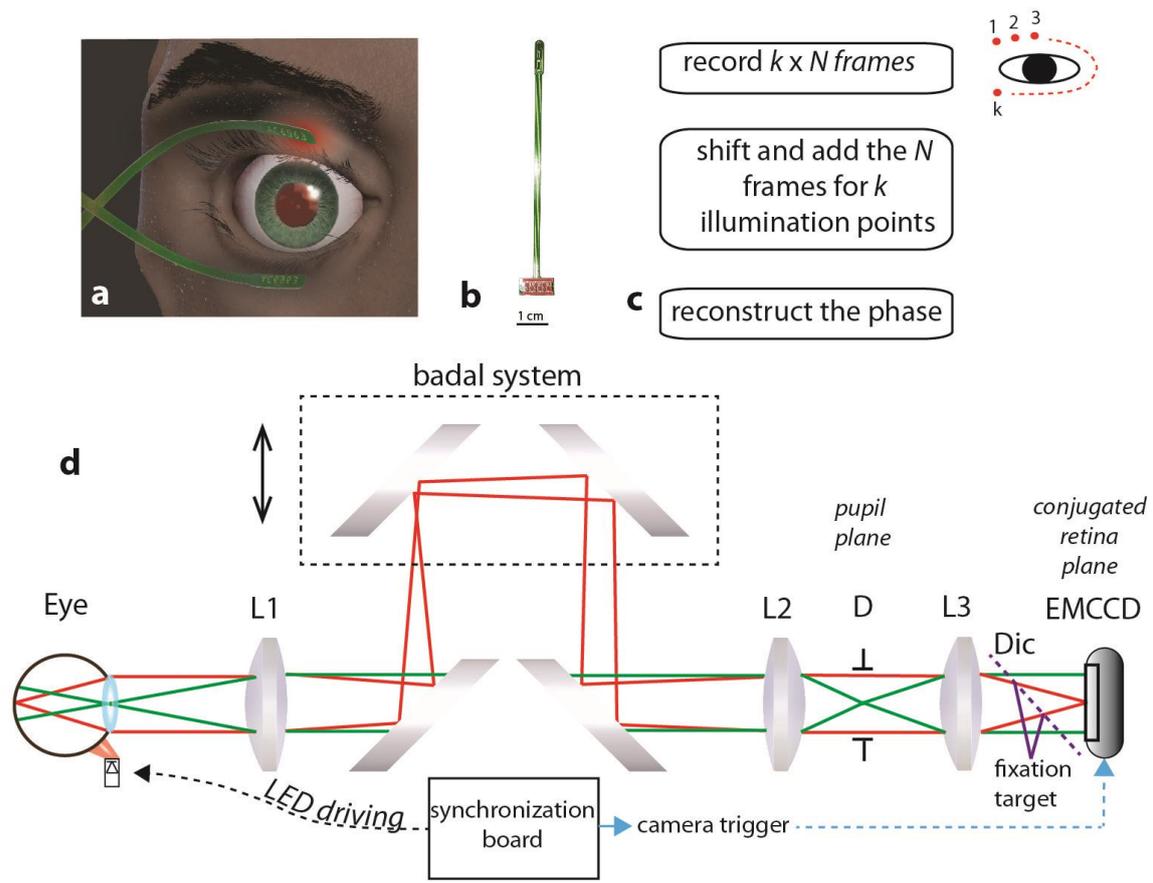

Figure 6: a. Trans-epidermal illumination by means of flexible PCB containing LEDs placed in contact with the skin of the eyelid. Light is then transmitted inside the eyeball. After scattering off the eye fundus, the light passing through the retinal's cell layers is collected by the eye lens. b. Flexible PCB holding 4 red LEDs. c. Recording and reconstruction procedure for in-vivo measurement. d. Experimental setup. The light scattered from the retina is collected by lens L1. The 4f system composed of the lenses L1 and L2 is adjusted for defocus thanks to a badal system. The lens L2 forms an image of the pupil plane at its focal distance, while the lens L3 forms an image of the retina on the EMCCD camera. Dic: dichroic mirror. Synchronization between the LEDs and the camera is performed thanks to a programmable board.

Figure 6c illustrates the procedure for recording and computing the images. It is as follows: 1) The flexible PCBs are placed on the patient's eyelids and maintained with adhesive tape. 2) the patient's head is placed on an ophthalmic head mount which holds the forehead and the chin. 3) Once the eye is aligned with the setup, the badal system is adjusted to focus on a retina layer. The sectionning is here limited by the pupil size and the aberrations. For instance, for an aberration free 6 mm pupil, the depth of field is  . 4) A sequence of a few hundred images is acquired by turning successively the LEDs ON and OFF with a ~100 ms exposure time. The procedure takes approximately 30s . 5) The images of each illumination point are averaged following the process described in [29]. 6) The phase image is reconstructed thanks to the regularization process of [16].

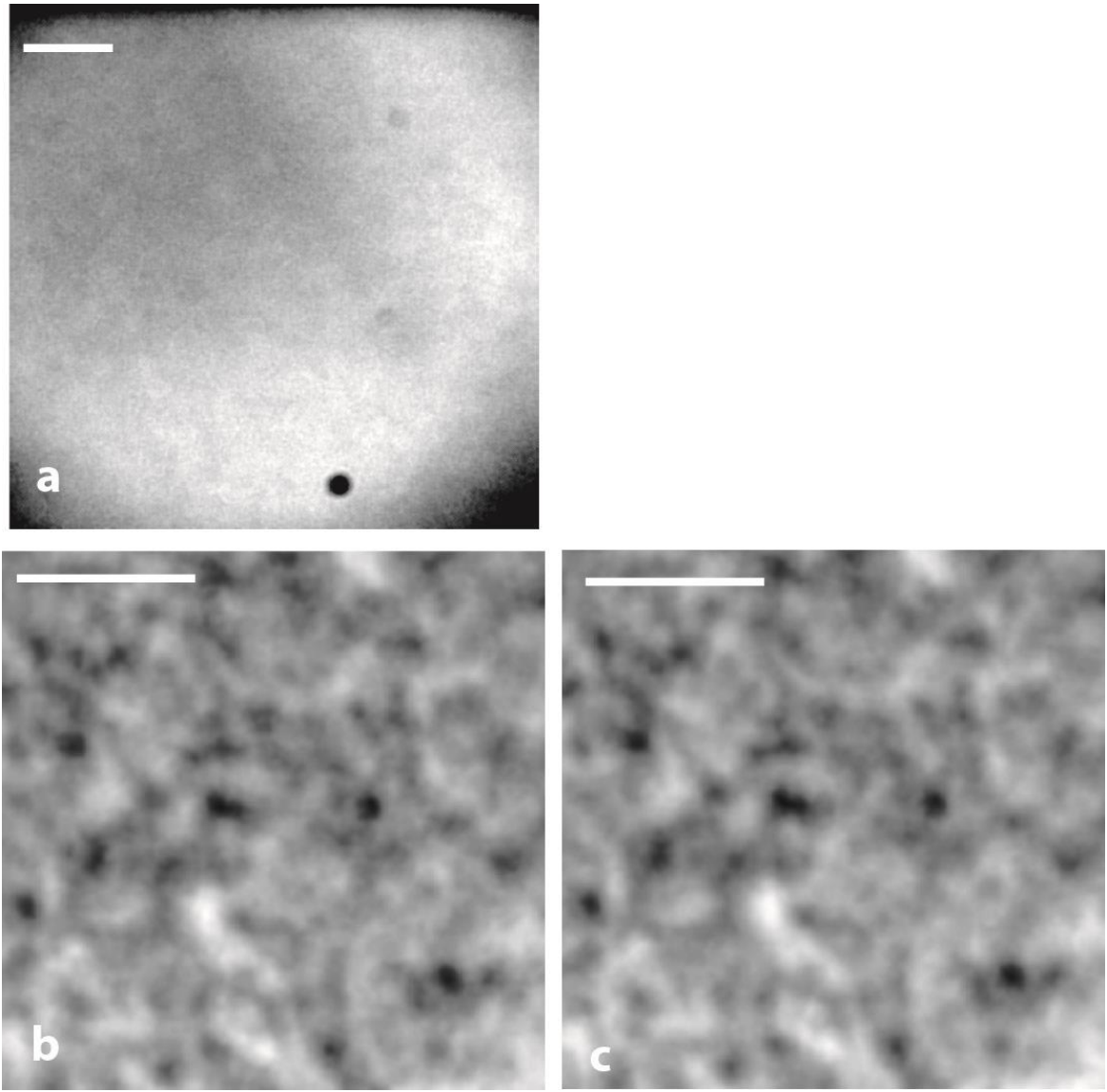

Figure 7: in vivo results at the fovea with 6 mm pupil. Raw dark field image a, shifted and added images from 2 illumination points b,c. scale bars = 200 µm.

The *in-vivo* images were acquired with a pupil of 6 mm, without dilation in dark room conditions. Illumination is provided with 850 nm peak wavelength LEDs, a radiant intensity of 1.85 mW/sr and a view angle of 145°. This gives a maximum ouput power of 8.03 mW. For electrical purpose, we lower the power at 4mW. Since the LEDs are in contact with the skin (still protected by a transparent isolating tape), the light flux emitted by the LED illuminates the skin over the area of the LED which is 1.2 $mm^2$. Based on European standard, the maximum permissible exposure of the skin for wavelength between 400 nm and 1400 nm and an exposure duration of 0.2 s, is 37000 W/$mm^2$ or 4.4 mW for 1.2 $mm^2$. For the retina, the maximum permissible exposure is given for a wavelength between 700 nm and 1050 nm, and exposure duration higher than 10 s. The MPE is then 75 µW/$mm^2$. Given an optical transmission of the skin of 70% [33], 40% for the sclera [34, 35], 1.12 mW are illuminating the retina over the scattered area. The illuminated area of the retina can be calculated by geometrical considerations, given a thickness of the skin and the sclera of 3 mm and the view angle of the LED of 145°. This leads to a power of 14 µW/$mm^2$. After passing once through thr retina on the side of the eyeball, the scattered light reaches the fundus on an even larger area.

As a comparison, the work presented in [32] reports a light power at the cornea of approximately 290 µW using a scanning laser system. The 290 µW are then focused on the retina in order to obtain one pixel on then reconstructed images. For in vivo human measurements 6 to 8 frames were averaged [32], given a value of power per point between 1.7 mW and 2.3 mW. In our study, we took advantage of having a single shot image recording that does not require adaptive optive to observe neurons. We simply averaged a hundred images to obtain a dark field image. With a PSF diameter of 2 µm (maximum resolution at NIR for aberration free eye), our experiment deals with 6 nW per point. In addition, the angle of illumination is larger than the maximum angle of the offset aperture, collected through the pupil. This provides higher contrast of the phase objects in the retina.

Fig 7 shows the in vivo results. The images were recorded in dark room condition, with non-dilated pupil. Indeed, the pupil diameter is approximately 6 mm. Fig 7a is a raw image acquired with 2 LEDs to create one illumination point, and an exposure time of 150 ms. The images is dominated by the noise but some features are clearly visible. Fig 7b and 7c correspond to the averaged images after shift and add, using an oversampling factor of 2, from 2 illumination points having 2 LEDs each.

**Conclusion**

We present a new microscopy for phase retinal imaging at cellular resolution. Our method takes advantage of the transcleral illumination obtaining high angle of illumination and so improving the contrast of the dark field image. The single shot camera system we developed is furthermore robust against noise and eye motions. It also prevents from using hardware based adaptive optics system, thanks to computational aberration correction. We first validated our method in vitro through a characterization of pig retina, performing cells density measurements. Finally, we demonstrated in vivo human imaging of the retina with a custom device implementing our modality. The ability of observing neuronal structures of the retina leads to great opportunities to improve the understanding of the eye pathologies.

**Sample preparation**.

Pigs eye were obtained from a local slaughterhouse, a few house after the death the eye were fixed with 4% paraformaldehyde for 24 h. Next, the eyes were stored 1 day in PBS before mounting. Flat mounting was performed detaching the choroid and retina from the sclera, with a few drops of PBS with glycerol solution.

For the human eye, the same protocol was applied, with flat mounting of the choroid and retina with PBS+glycerol solution.

**References**

[1] R. H. Masland, *The Neuronal Organization of the Retina* Neuron. 2012 October 18; 76(2): 266–280.

[2] Fujimoto JG, Bouma B, Tearney GJ, Boppart SA, Pitris C, Southern JF, Brezinski ME.New technology for high-speed and high-resolution optical coherence tomography.Ann N Y Acad Sci. 1998 Feb 9;838:95-107.

[3] Drexler W, Fujimoto JG. State-of-the-art retinal optical coherence tomography. Prog Retin Eye Res. 2008 Jan;27(1):45-88.

[4] Jonnal RS, Kocaoglu OP, Zawadzki RJ, Liu Z, Miller DT, Werner JS. A Review of Adaptive Optics Optical Coherence Tomography: Technical Advances, Scientific Applications, and the Future. Invest Ophthalmol Vis Sci. 2016 Jul 1;57(9):OCT51-68

[5] Drew Scoles, Yusufu N. Sulai, and Alfredo Dubra, *In vivo dark-field imaging of the retinal pigment epithelium cell mosaic*, Biomed. Opt. Express **4**, 1710-1723 (2013).

[6] Liu Z, Kocaoglu OP, Miller DT. *3D imaging of retinal pigment epithelial cells in the living human retina*. Invest Ophthalmol Vis Sci. 2016, 57:OCT533–OCT543.

[7] Toco Y. P. Chui, Dean A. VanNasdale, and Stephen A. Burns, *The use of forward scatter to improve retinal vascular imaging with an adaptive optics scanning laser ophthalmoscope*, Biomed. Opt. Express **3**, 2537-2549 (2012)

[8] Toco Y. P. Chui, Thomas J. Gast, Stephen A. Burns; *Imaging of Vascular Wall Fine Structure in the Human Retina Using Adaptive Optics Scanning Laser Ophthalmoscopy*. *Invest. Ophthalmol. Vis. Sci.* 2013;54(10):7115-7124.

[9] W. B. Amos, S. Reichelt, D. M. Cattermole, and J. Laufer, "Re-evaluation of differential phase contrast (DPC) in a scanning laser microscope using a split detector as an alternative to differential interference contrast (DIC) optics," J. Microsc. 210(2), 166–175 (2003).

[10] David Cunefare, Robert F. Cooper, Brian Higgins, David F. Katz, Alfredo Dubra, Joseph Carroll, and Sina Farsiu, *Automatic detection of cone photoreceptors in split detector adaptive optics scanning light ophthalmoscope images*, Biomed. Opt. Express **7**, 2036-2050 (2016)

[11] Y. N. Sulai, D. Scoles, Z. Harvey and A. Dubra, *Visualization of retinal vascular structure and perfusion with a nonconfocal adaptive optics scanning light ophthalmoscope*, Vol. 31, No. 3 March 2014, J. Opt. Soc. Am. A.

[12] A. Guevara-Torres, D. R. Williams, and J. B. Schallek, *Imaging translucent cell bodies in the living mouse retina without contrast agents*, Biomed. Opt. Express **6**, 2106-2119 (2015).

[13] T. N. Ford, K. K. Chu, and J. Mertz, *Phase-gradient microscopy in thick tissue with oblique back-illumination*, Nat. Methods 9, 1195–1197 (2012).

[14] J. David Giese, Tim N. Ford, and Jerome Mertz, *Fast volumetric phase-gradient imaging in thick samples*, Opt. Express **22**, 1152-1162 (2014).

[15] Lei Tian and Laura Waller, *Quantitative differential phase contrast imaging in an LED array microscope*, Opt. Express 23, 11394-11403 (2015).

[16] S. B. Mehta and C. J. Sheppard, *Quantitative phase-gradient imaging at high resolution with asymmetric illumination-based differential phase contrast*, Opt. Lett. 34, 1924–1926 (2009).

[17] L. Tian, J. Wang, and L. Waller, *3D differential phase-contrast microscopy with computational illumination using an LED array*, Opt. Lett. 39, 1326–1329 (2014).

[18] Z. Liu, L. Tian, S. Liu, and L. Waller, Real-time brightfield, darkfield, and phase contrast imaging in a light-emitting diode array microscope, J. Biomed. Opt. 19, 106002 (2014).

[19] F. Zernike, Physica 9, 974 (1942).

[20] G. Nomarski, J. Phys. Radium 16, 9 (1955).

[21] P. Marquet, B. Rappaz, P. J. Magistretti, E. Cuche, Y. Emery, T. Colomb, and C. Depeursinge, *Digital holographic microscopy: a noninvasive contrast imaging technique allowing quantitative visualization of living cells with subwavelength axial accuracy*, Opt. Lett. 30, 468-470 (2005)

[22] D T Miller, O P Kocaoglu, Q Wang and S Lee,*Adaptive optics and the eye (super resolution OCT)*, Eye (2011) 25, 321–330.

[23] N. D. Shemonski ,FredrickA.South, Yuan-Zhi Liu,Steven G.Adie,P.ScottCarney and S. A. Boppart *Computational high-resolution optical imaging of the living human retina* Nat. Photonics 9, 440–443 (2015).

[24] D. Hillmann, H. Spahr, C. Hain, H. Sudkamp, G. Franke, C. Pfäffle, C.Winter, and G. Hüttmann *Aberration-free volumetric high-speed imaging of invivo retina*. arXiv, May, 2016.

[25] R. S. Jonnal, O. P. Kocaoglu, R. J. Zawadzki, Z. Liu, D. T. Miller, J. S. Werner, *A Review of Adaptive Optics Optical Coherence Tomography: Technical Advances, Scientific Applications, and the Future*. *Invest. Ophthalmol. Vis. Sci.* 2016;57(9):OCT51-OCT68.

[26] Garcia et al, Topography of pig retinal ganglion cells, J Comp Neurol. 2005 Jun 13;486(4):361-72.

[27] Chandler et al, Photoreceptor density of the domestic pig retina, Vet Ophthalmol. 1999;2(3):179-184.

[28] Gerke et al, Topography of rods and cones in the retina of the domestic pig, Hong Kong Med J 1995;1:302-8.

[29] N. Meitav and E. N. Ribak, *Improving retinal image resolution with iterative weighted shift-and-add*, J. Opt. Soc. Am. A **28**, 1395-1402 (2011).

[30] J. Kühn, E. Shaffer, J. Mena, B. Breton, J. Parent, B. Rappaz, M. Chambon, Y. Emery, P. Magistretti, C. Depeursinge, P. Marquet, G. Turcatti, *Label-Free Cytotoxicity Screening Assay by Digital Holographic Microscopy*. Assay and Drug Development Technologies, *11*(2), 101–107(2013).

[31] B. Rappaz, P. Marquet, E. Cuche, Y. Emery, C. Depeursinge, and P. J. Magistretti, *Measurement of the integral refractive index and dynamic cell morphometry of living cells with digital holographic microscopy*, Opt. Express **13**, 9361-9373 (2005)

[32] E. A. Rossi, C. E. Granger, R. Sharma, Q. Yang, K. Saito, C. Schwarz, S. Walters, K. Nozato, J. Zhang, T. Kawakami, W. Fischer, L. R. Latchney, J. J. Hunter, M. M. Chung and D. R. Williams, *Imaging individual neurons in the retinal ganglion cell layer of the living eye* PNAS 2017 114 (3) 586-591; published ahead of print January 3, 2017.

[33] S. L. Jacques *Optical properties of biological tissues: a review*, Physics in Medicine and Biology, Vol. 58, no 11 , 2013.

[34] H. Hammer, D. Schweitzer, E. Thamm, A. Kolb, J. Strobel *Scattering properties of the retina and the choroids determined from OCT-A-scans*, Int Ophthalmol. 2001;23(4-6):291-5.[35] A. Vogel, C. Dlugos , R. Nuffer , R. Birngruber,Optical properties of human sclera, and their consequences for transscleral laser applications. Lasers Surg Med. 1991;11(4):331-40.